\documentclass[prl,showpacs,preprintnumbers,amsmath,amssymb]{revtex4}

\usepackage{graphicx}
\usepackage{dcolumn}
\usepackage{bm}

\begin{document}

\title{\bf Hydrogen dissociation on the Mg(0001) surface from 
quantum Monte Carlo calculations}

\author{M. Pozzo$^{1,2}$}
\author{D. Alf\`{e}$^{1,2,3,4}$}\email{d.alfe@ucl.ac.uk}
\affiliation{
$^1$Department of Earth Sciences, UCL, Gower
Street, London WC1E 6BT, United Kingdom, \\
$^2$Materials Simulation Laboratory, UCL, Gower Street, 
London WC1E 6 BT, United Kingdom,\\
$^3$Department of Physics and Astronomy, UCL, Gower
Street, London WC1E 6BT, United Kingdom \\
$^4$London Centre for Nanotechnology, UCL, 17-19 Gordon
Street, London WC1H 0AH, United Kingdom }

\date{\today}

\begin{abstract}

We have used diffusion Monte Carlo (DMC) simulations to calculate the
energy barrier for H$_2$ dissociation on the Mg(0001) surface.  The
calculations employ pseudopotentials and systematically improvable
B-spline basis sets to expand the single particle orbitals used to
construct the trial wavefunctions. Extensive tests on system size,
time step, and other sources of errors, performed on periodically
repeated systems of up to 550 atoms, show that all these errors
together can be reduced to $\sim 0.03$ eV. The DMC dissociation
barrier is calculated to be $1.18 \pm 0.03$~eV, and is compared to
those obtained with density functional theory using various
exchange-correlation functionals, with values ranging between 0.44 and
1.07 eV.

\end{abstract}

\maketitle

\section{Introduction}

The ability of calculating accurately the energetics of metal hydrides
is a key issue in the design of novel materials, which might turn out
to be useful in the current search for suitable hydrogen storage (HS)
media. In particular, the kinetics of hydrogen absorption is one of
the main questions which determine the functionality of a candidate HS
material.

Magnesium is an interesting hydrogen absorbant, because the hydride
(MgH$_2$) formed after exposure to H$_2$ can store a large amount of H
(7.6 \% by weight), which can be later released by heating the
material above $\sim 300~^o$C\cite{bogdanovic99}, the reaction being
endothermic with an enthalpy of decomposition of 76
kJ/mole~\cite{yamaguchi94}. The kinetics of hydrogen intake by Mg is
quite slow, because of a relatively large energy barrier that the
H$_2$ molecule needs to overcome in order to dissociate before
adsorption.  However, it is possible to modify this material to
improve its properties, for example by doping it with traces of
transition
metals~\cite{liang99,vajeeston02,yavari03,liang04,shang04,song04,hanada05,vansetten05,vegge05},
which have been shown to be very effective at reducing the activation
energy for hydrogen dissociation
~\cite{vegge04,du05,pozzo08a,pozzo08b}, and also to somewhat reduce
the decomposition temperature of the hydride~\cite{shang04}.

So far, the only high level quantum mechanics calculations available
for the interaction of hydrogen with Mg surfaces are based on density
functional theory (DFT), which has been shown to be quite successful 
for investigating the relative energetics of the interaction of H$_2$ 
with pure and transition metal doped Mg
surfaces~\cite{vegge04,du05,pozzo08a,pozzo08b}. 
However, the absolute accuracy of DFT has often been called into question. 
For example, we recently showed that the DFT calculated value for the
enthalpy of formation of MgH$_2$ from Mg bulk and H$_2$, a quantity
obviously related to the interaction between H and Mg, strongly
depends on the functional employed, with values ranging from 0.29 eV (with RPBE)
to 0.82 eV (with LDA)~\cite{pozzo08c}.

Quantum Monte Carlo (QMC) methods~\cite{foulkes01,umrigar93} are
becoming promising techniques to improve beyond DFT, 
with their increased utilisation favoured by the
availability of faster and faster computers.  Recently, diffusion
Monte Carlo (DMC) calculations on Mg and MgH$_2$ bulk showed very good
accuracy in both structural parameters, cohesive energies, and the
enthalpy of formation of the MgH$_2$ crystal~\cite{pozzo08c}.

\begin{figure}
\centerline{
\includegraphics[width=2.5in,angle=-90]{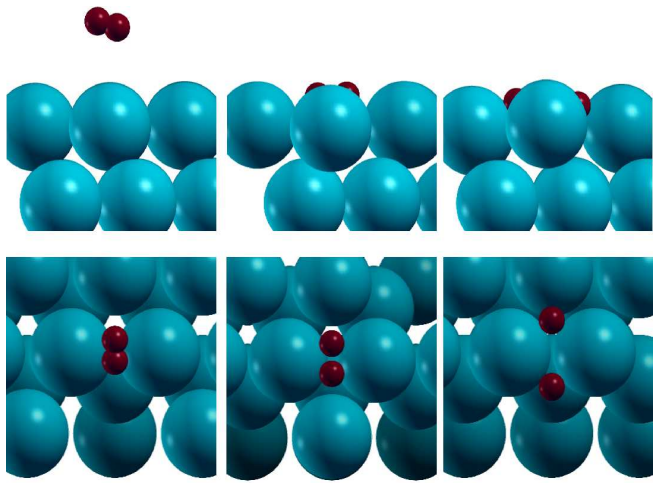}
}
\caption{(Color online) Initial state (left), transition state
(centre) and final state (right) geometries for the H$_2$ dissociation
reaction on the Mg(0001) surface. Top and bottom panels show side and
top view respectively. }\label{fig:is_ts_fs}
\end{figure}

Here we have coupled DMC to climbing image nudget elastic band (NEB)
DFT calculations to compute the energy barrier for the dissociation of
H$_2$ on the Mg(0001) surface. Since DMC forces are not yet readily
available, the strategy has been to first perform DFT calculations to
determine the geometries of the problem: i) the initial state (IS)
with the H$_2$ molecule in vacuum plus the clean Mg(0001) surface, ii)
the final state (FS) with the two H atoms adsorbed on two nearby
hollow sites, and iii) the transition state (TS) determined with the
NEB method (see Fig.~\ref{fig:is_ts_fs}); then, we have used these DFT
geometries to compute highly accurate DMC energies.

\section{Technical details}

DFT calculations have been performed with the {\sc vasp}
code~\cite{kresse96}, using the projector augmented method
(PAW)~\cite{blochl94,kresse99} with the generalised gradient
approximations known as PBE~\cite{pbe}, PW91~\cite{pw91},
RPBE~\cite{rpbe} or the local density approximation (LDA).  An
efficient charge density extrapolation was used to speed up the
calculations~\cite{alfe99}. The plane-wave (PW) energy cut-off was 270
eV, which guarantees convergence of adsorption energies within 1
meV~\cite{pozzo08a}.  Surfaces were modelled using periodic slabs,
with 5 atomic layers and a vacuum thickness of 10~\AA. The topmost
three atomic layers were allowed to relax, while the bottom two were
held fixed to the positions of bulk Mg. The $c/a$ ratio and the
lattice parameter were fixed at 1.621 and 3.2~\AA\, respectively,
which are close to the calculated PBE values at zero pressure and zero
temperature~\cite{pozzo08a}.  We used 2x2 surface unit cells (22
atoms), with 9x9x1 {\bf k}-point grids. These settings were
extensively tested (repeating calculations on 3x3 surface unit cells
and, separately, denser {\bf k}-point grids) and guarantee convergence
of activation energies to better than 0.02 eV~\cite{pozzo08a,valence}.
Activation energies have been calculated with the NEB
method~\cite{neb} using 9 replicas, and tested with calculations using
17 replicas; 9 replicas proved to be sufficient to reach convergence
of activation energies to better than 0.01 eV and display all the main
features of the minimum energy path.

DMC calculations have been performed using the {\sc casino}
code~\cite{casino}, employing trial wavefunctions of the
Slater-Jastrow type: $\Psi_T ( {\bf R} ) = D^\uparrow D^\downarrow
e^{J}$, where $D^\uparrow$ and $D^\downarrow$ are Slater determinants
of up- and down-spin single-electron orbitals. The Jastrow factor
$e^J$ is the exponential of a sum of one-body (electron-nucleus),
two-body (electron-electron) and three body
(electron-electron-nucleus) terms, which are parametrised functions of
electron-nucleus, electron-electron and electron-electron-nucleus
separations, and are designed to satisfy the cusp conditions. The
parameters in the Jastrow factor are varied to minimise the variance
of the local energy $E_L ( {\bf R} ) \equiv \Psi_T^{-1}({\bf
  R})\hat{H} \Psi_T({\bf R})$.  Imaginary time evolution of the
Schr\"odinger equation has been performed with the usual short time
approximation, and the locality approximation~\cite{mitas91}. We
extensively tested time step errors in our previous work on
MgH$_2$~\cite{pozzo08c}, and used here a time step of 0.05 a.u., which
results in essentially negligible energy biases.  We used Dirac-Fock
pseudopotentials (PP) for Mg and H~\cite{trail05}. The Mg PP has a
frozen Ne core and a core radius of 1.43~\AA, the H PP has a core
radius of 0.26~\AA.  The single particle orbitals have been obtained
by DFT-PW calculations with the the LDA and a PW cutoff of 3400 eV,
using the {\sc pwscf} package~\cite{pwscf}, and re-expanded in terms
of B-splines~\cite{hernandez97,alfe04}, using the natural B-spline
grid spacing given by $ a = \pi / G_{\rm max} $, where $G_{\rm max}$
is the length of the largest PW.  The DMC calculations were performed
using the Ewald technique to model electron-electron interactions.
The number of walkers in the DMC simulations varied between 2560 and
10240, depending on the size of the system.

\section{Results}

\begin{figure}
\centerline{
\includegraphics[width=3.4in]{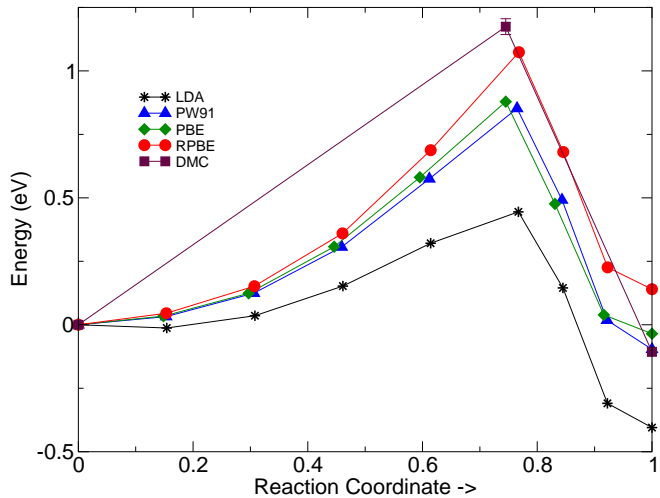}
}
\caption{(Color online) Minimum energy profiles for the H$_2$
dissociation reaction on the Mg(0001) surface calculted with various
DFT functionals: LDA (black stars), PW91 (blue triangles), PBE (green
diamonds) and RPBE (red circles). Also shown are the DMC energies
(maroon squares) at the TS and the FS. Lines are guides to the
eye.}\label{fig:barriers}
\end{figure}

\begin{table}
\caption{\label{table1} Average distances of molecular hydrogen from
the first surface layer at the transition state ($\overline{d}^{\,TS}_{H-surf}$)
and final state ($\overline{d}^{\,FS}_{H-surf}$) of the minimum energy
paths calculated with various exchange-correlation functionals. Also
reported are the distances between the hydrogens at the transition state ($d^{\,TS}_{H-H}$) 
and the activation barrier for H$_2$ dissociation
(E$_{TS-IS}$) and the energy difference between the final and initial
state (E$_{FS-IS}$) obtained with the four DFT functionals and with
DMC. }
\begin{ruledtabular}
\begin{tabular}{lccccc}
PP & $\overline{d}^{\,TS}_{H-surf}$ (\AA) & $d^{\,TS}_{H-H}$ & $\overline{d}^{\,FS}_{H-surf}$ (\AA) & E$_{TS-IS}$ (eV) & E$_{FS-IS}$ (eV) \\ 
\hline
LDA  & 1.09 & 1.02 & 0.74 & 0.44 & -0.40 \\
PW91 & 1.07 & 1.07 & 0.79 & 0.85 & -0.10 \\
PBE  & 1.07 & 1.08 & 0.81 & 0.88 & -0.03 \\
RPBE & 1.07 & 1.0  & 0.81 & 1.07 & 0.14 \\
DMC &     &   & 1.18$\pm$0.03 & -0.11$\pm$0.02 \\
\end{tabular}
\end{ruledtabular}
\end{table}

DFT energy barriers were calculated using 4 different functionals:
LDA, PW91, PBE and RPBE. The energy profiles along the minimum energy
paths are displayed in Fig.~\ref{fig:barriers}. We also calculated the
LDA, PW91 and RPBE energies at the PBE IS, TS and FS. In all cases the
differences between these energies and the corresponding energies
obtained in the LDA, PW91 and RPBE IS, TS and FS are between 1 and 2
meV. This also indicates that the geometries obtained with the 4
different functionals very similar, and indeed this is confirmed by
looking at the distances of the hydrogen atoms from the surface at the
TS and at the FS (see Table~\ref{table1}). In Table~\ref{table1} we
also report the energy barriers and the energy differences between the
FS and the IS calculated with the 4 functionals and with DMC (see
below). Our DFT-RPBE atcivation energies compare well with the results
of Du et al.~\cite{du05} and Vegge~\cite{vegge04}.

DMC calculations were performed using the PBE geometries for the IS,
the TS and the FS, as shown in Fig.~\ref{fig:is_ts_fs}.  We used 2x2
(88 atoms), 3x3 (198 atoms), 4x4 (352 atoms) and 5x5 (550 atoms)
periodically repeated supercells of the 22 atoms cell used to perform
the DFT calculations. These large calculations were performed to
investigate DMC finite size errors.  Raw DMC energies were corrected
using the difference between the LDA {\bf k}-point converged energies
and the LDA energies with a {\bf k}-point sampling corresponding to
the same simulation cell employed in the DMC calculations. These
corrections make it easier to extrapolate the DMC results to infinite
size.

\begin{figure}
\centerline{
\includegraphics[width=3.4in]{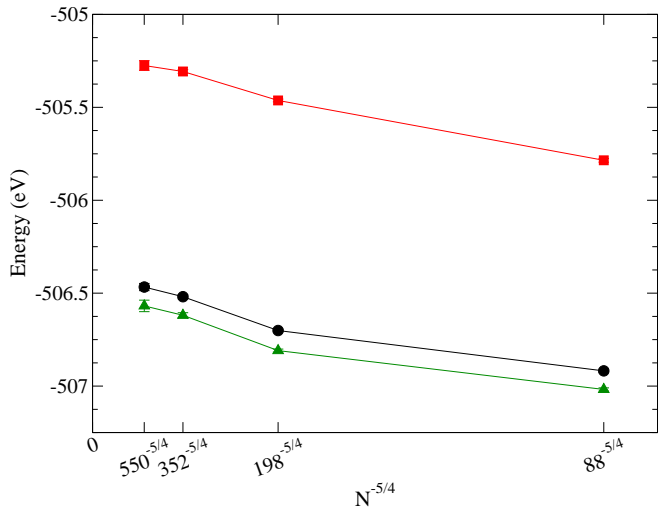}
}
\caption{(Color online) Diffusion Monte Carlo energies for the initial
state (black circles), the transition state (red squares), and the
final state (green triangles) for the H$_2$ dissociation reaction on
the Mg(0001) surface as function of $N^{-5/4}$, with $N$ the number
of atoms in the simulation cell.}\label{fig:size}
\end{figure}

\begin{figure}
\centerline{
\includegraphics[width=3.4in]{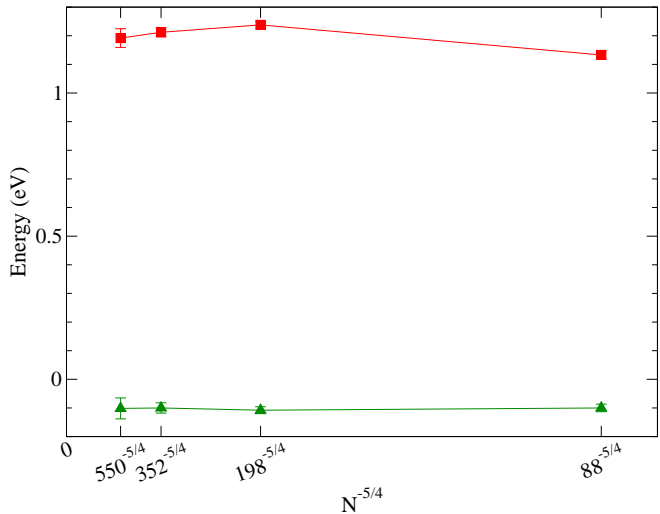}
}
\caption{(Color online) Diffusion Monte Carlo energy differences
between the transition state and the initial state (red squares) and
between the final state and the initial state (green triangles) for
the H$_2$ dissociation reaction on the Mg(0001) surface as function
of $N^{-5/4}$, with $N$ the number of atoms in the simulation
cell. }\label{fig:diff}
\end{figure}

It was shown by Drummond et al.~\cite{drummond08} that for
calculations on 2-dimensional systems DMC energies obtained using the
Ewald interaction should have $N^{-5/4}$ behaviour, where $N$ is the
number of atoms (or, more generally, electrons) in the system. They
showed that the leading error in the kinetic energy of the 2D
homogeneous electron gas falls off as $o(N^{-5/4})$. The system
studied here is a 2D slab, and we therefore follow their
recommendation of assuming an error in the total energies of the slab
going as $N^{-5/4}$.  In Fig.~\ref{fig:size} we show the DMC energies
for the IS, the TS and the FS as function of $N^{-5/4}$, and we notice
that if we disregard the smallest system (88 atoms) the other three
sizes fit more or less on a straight line. In fact, the behaviour is
not exactly linear, but the deviations appear to be similar for the
IS, the TS and the FS. In Fig.~\ref{fig:diff} we plot the energy
differences between the TS and the IS, and between the FS and the IS
as function of $N^{-5/4}$. The finite size dependence of these
differences are much weaker, and easily extrapolable to infinite size,
if just the smallest system (88 atoms) is not included in the fit.

The DMC extrapolated results to infinite size are plotted in
Fig.~\ref{fig:barriers}, from which we observe that at the TS the RPBE
result appears to be the nearest, while at the FS the PBE and the PW91
functionals appear to be those with the closest energy. 

\section{Discussion and Conclusions}

Before discussing our findings, we should point out once more that the
present DMC results for the dissociation barrier are not the ``real''
DMC results for the activation barrier, because we used DFT-PBE
geometries for the IS the TS and the FS. At present this is all we can
do, because we do not have easy access to DMC forces, however, we
argued that the similarity of the geometries found with the 4 DFT
functional emplyed here may suggest that these geometries are indeed
reliable. 

The present results would suggest that RPBE is the best functional to
calculate the dissociation energy barrier, however, for the reverse
barrier this would not be the case. Other related quantities are also
badly calculated with RPBE, like the enthalpy of formation of MgH$_2$
from Mg bulk and H$_2$ that we find to be 0.29 eV (this includes zero
point energies), which is about 0.5 eV lower than the experimental
value of 0.79 eV, or the DMC value of 0.85 eV~\cite{pozzo08c}.  As far
as this enthalpy of formation is concerned, the RPBE is the functional
which has the worse agreement with DMC among LDA, PBE, PW91 and RPBE.

We found in the literature two experimental values for dissociation
energy barrier.  The first one reported in Ref.~\cite{sprunplu91}
($\sim$ 1.0 eV) refers to the recombination barrier (which, in this
particular case, is similar to the dissociation barrier) identified
with the barrier for desorption from the surface.  This value was not
directly measured in the thermal programmed desorption (TPD)
experiments of Ref.~\cite{sprunplu91} because complete desorption
spectra as function of temperature could not be taken, due to the
onset of Mg sublimation at $\sim$ 450 K, which overlaps with the
temperature at which H$_2$ desorbs. However, it was noted that the
onset of H$_2$ desorption appears at 425 K, which is similar to that
of the H/Be(0001) system that has a determined desorption energy of
$\sim$ 1 eV~\cite{ray90}, and so, by analogy, it was suggested that
the activation energy for desorption might be the same on the
H/Mg(0001) system too.

The second, and most recent one, refers to the dissociation of H$_2$
on a 400~\AA~thick magnesium film, and has a reported value for the
dissociation barrier of 0.75$\pm$0.15 eV~\cite{johansson06}.  This
value is in good agreement with the calculated PBE dissociation
energy, but is significantly lower than the present DMC dissociation
energy, even if a correction of ~0.08 eV is applied to take zero point
effects into account (see Vegge~\cite{vegge04}).  It should be noted,
however, as pointed out in Ref.~\cite{johansson06}, that the
experimental situation may not be the same as the theoretical ones,
due to the possible presence of steps on the surface which might be
more reactive sites and lower the H$_2$ dissociation barrier. We
investigated possible effects due to the presence of steps, and we
found that the energy barrier is hardly affected by steps on the
Mg(0001) surface, so this might not be the reason for the apperent
higher activity inferred from these experimental data. We speculate
that a more likely reason might be the following.  The experimentally
inferred dissociation energy of 0.75$\pm$0.15 eV is based on the use
of the Arrhenius relations with assumed pre-factors of $\sim
10^{12}$~Hz.  As showed in Refs.~\cite{alfe06} and~\cite{alfe07},
these values could be underestimated by more than two orders of
magnitudes because the classical prefactors do not include the
enhancement due to the much larger entropy that the molecules have in
vacuum. It is possible that just these prefactor enhancement could
explain an activation energy up to $\sim 0.25$ eV higher, or in other
words that the correct experimental number may, in fact, be
1.0$\pm$0.15 eV, which would agree well with the present DMC results.
It will be interesting to investigate this possiblity to confirm or
otherwise this suggestion, and we plan to do this in the future.

In summary, we have performed here DFT and DMC calculations for the
activation energy needed to dissociate the H$_2$ molecule on the
Mg(0001) surface. We have shown that the four functionals employed,
LDA, PW91, PBE and RPBE, all give different results, with values
ranging from 0.44 to 1.07 eV. Such a large variability calls into
question the reliability of certain DFT functionals, and makes it difficult to make
choices on the functional to employ when highly accurate energy values
are needed.  However, we showed that the geometries of the IS, TS and
FS are predicted to be very similar by the 4 functionals, which
therefore provides some confidence that these geometries are indeed
accurate. We have then used these geometries (the PBE ones, to be
precise), to calculate DMC energies, and obtained a value for the
activation energy for H$_2$ dissociation of $1.18 \pm 0.03$~eV, which
is close to the DFT-RPBE value of 1.07 eV. The inverse barrier is
calculated to be $1.28 \pm 0.03$~eV, which is at least 0.3 eV higher
than any of the results obtained with the 4 DFT functionals, and casts
doubts on the reliability of DFT adsorption energies.

\section*{Acknowledgements}
The computations were performed on the HECToR service.  This work was
conducted as part of a EURYI scheme as provided by EPSRC (see
www.esf.org/euryi).


\begin{thebibliography}{99}
\bibitem{valence} We used PAW potentials with only the $3s^2$ electrons in valence.
Tests including also the $2p^6$ showed differences of less than 0.02 eV in energy barriers
and adsorption enegies.
\end{thebibliography}
\end{document}